\documentstyle[12pt,epsf]{article}
\newcommand{\tdm}[1]{\mbox{\boldmath $#1$}}

\newcommand{\gev}{\rm \; GeV}

\begin{document}
\begin{titlepage}
\pagestyle{empty}
\vspace*{1cm}
\begin{center}
{\large\bf
The QCD pomeron at TESLA --- motivation and exclusive $J/\psi$ production 
\footnote{
To appear in the Proceedings of the 2nd Joint ECFA/DESY Study on Physics and
Detectors for a Linear Electron-Positron Collider, Obernai, France, 
October 1999.}
}
\vspace{1.1cm}\\
         {\sc J.~Kwieci\'nski}$^a$, 
         {\sc L.~Motyka}$^b$ and
         {\sc A. De Roeck}$^c$
\vspace{0.3cm}\\
$^a${\it Department of Theoretical Physics, \\
H.~Niewodnicza\'nski Institute of Nuclear Physics,
Cracow, Poland}
\vspace{0.3cm}\\
$^b${\it Institute of Physics, Jagellonian University,
Cracow, Poland}\\[0.3cm]
$^c${\it CERN, Geneva, Switzerland}
\end{center}
\vspace{1.5cm}
\begin{abstract}
We briefly present the motivation for studying the processes
mediated by the QCD pomeron at high energy $e^+e^-$ colliders.
We describe the behaviour of the cross-section for
the reaction $\gamma\gamma \to J/\psi J/\psi$ obtained
from the BFKL equation with dominant non-leading corrections.
We give the predictions for the rates of anti-tagged
$e^+e^- \to e^+e^- J/\psi J/\psi$ events in TESLA and
conclude that such  reactions may be excellent probes
of the hard pomeron.
\end{abstract}
\vspace{1cm}

\noindent
\end{titlepage}

The pomeron exchange is one of the most intriguing phenomena in quantum
chromodynamics. It governs the leading behaviour of the scattering
amplitude for two objects which may interact by exchanging colour degrees
of freedom. The characteristic feature of these amplitudes is an
approximate power-like dependence on the collision energy, which supports
the picture of the pomeron as an isolated Regge pole in the complex angular
momentum plane. The exponent characterizing the pomeron exchange is
universal for soft processes and thus it is convenient to use
phenomenological Regge motivated models \cite{DL} for the pomeron. For hard
processes at high energies it is possible to go beyond the level of models
and describe the pomeron in the framework of perturbative QCD in the high
energy limit \cite{GLR,LIPAT1}.
However, at every order of the perturbative expansion there
appear large logarithms of the collision energy which accompany the
coupling constant and one cannot rely on the fixed order perturbation
theory. The necessary resummation of ladder diagrams with reggeized gluons
along the chain may be performed with the use of the Balitzkij, Fadin,
Kuraev, Lipatov (BFKL) equation \cite{BFKL1}. Formally, this equation gives
the exact answer in the leading logarithmic approximation, but the recently
obtained NLO corrections \cite{BFKLNL} are very large, that could make
the BFKL approach not useful for  phenomenology.
One is also forced to face the problem of contributions to the amplitudes
coming from the infra-red domain. In fact one can control both effects
in an approximate way and construct reasonable phenomenology on the
basis of the BFKL equation. The important role of the experiments is therefore
to verify the quality of the approximations and provide the guideline
for further development of the pomeron theory.

Unfortunately, for most observables, it is difficult to disentangle
the genuine hard pomeron effects from other contributions.
The ideal high energy process should have the following properties: \\
(i) The virtualities of the  gluons along the
ladder should be large enough to assure the relevance of the
perturbative expansion. The necessary hard scale may be provided either
by the coupling of the ladder to scattering particles, that contain a hard
scale themselves, or by large momentum transfer carried by the gluons. \\
(ii) In order to distinguish the genuine BFKL from DGLAP evolution effects
it is convenient to focus on processes in which the scales on the both
ends of the ladder are of comparable size. Then, the amplitude
is free from the DGLAP corrections due to zero length of the
DGLAP evolution.\\
(iii) Finally, one requires the non-perturbative effects to factor out.\\
Since these requirements are rather stringent, the list of
possible measurements in hadron-hadron and lepton-hadron colliders
is limited. It would be therefore desirable to study  pomeron
physics also for the two (virtual) photon processes at TESLA. The quality
of such measurements is expected to be excellent due to very
large luminosity for high energy photon-photon collisions and the clean
experimental signatures. In what follows we shall present the
estimates of the cross-section of the double exclusive $J/\psi$
production at TESLA which have been performed using the BFKL
equation with non-leading corrections \cite{KMPSI}. A typical
diagram illustrating this process is given in Fig.~1.

\noindent
\begin{figure}
\leavevmode
\begin{center}
\parbox{5.5cm}{
\epsfxsize = 5.0cm
\epsfysize = 4.2cm
\epsfbox{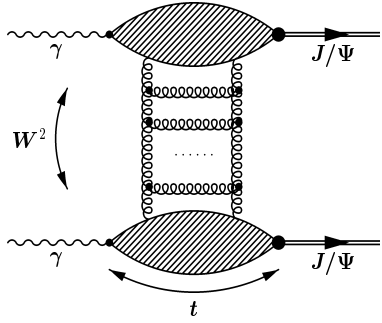}}
\end{center}
\caption{\small The QCD pomeron exchange mechanism of the processes
         $\gamma\gamma \rightarrow J/\psi J/\psi$.}
\end{figure}

The doubly exclusive $J/\psi$ productions is an unique process since
it allows to test the differential cross-section for arbitrary
momentum transfer. Besides that, the pomeron exchange amplitude
enters the cross-section in the second power making the dependence
on the energy very robust. The hard scale is given by the relatively
large mass of the $c$-quark and thus the perturbative calculation is
valid even for real photons or photons with small virtuality.
The flux of such photons in an electron contains a large enhancement
proportional to the logarithm of the beam energy which moves this
rather exotic process into the measurable domain. Since the
$e^+ e^-$ cross-section for $J/\psi$ production is dominated by
small virtualities of the mediating photons it is convenient
to focus on the anti-tagged $e^+e^-$ events.
The cross-section for the process $e^+e^- \rightarrow e^+e^- + Y$
for anti-tagged $e^{\pm}$ corresponds to the production of the hadronic
state~$Y$ in $\gamma\gamma$ collision and is given by a standard
convolution integral \cite{GGREV}.

The imaginary part ${\rm Im} A(W^2,t=-Q_P ^2)$ of the amplitude for
the considered process which corresponds to the diagram in Fig.~1
can be written in the following form:
\begin{equation}
{\rm Im} A(W^2,t=- Q_P^2) =
\int {d^2\tdm k \over\pi} {\Phi_0(k^2, Q_P ^2)\Phi(x,\tdm k,\tdm Q_P)\over
[(\tdm k + \tdm Q_P /2)^2 +s_0][(\tdm k - \tdm Q_P /2)^2+s_0]}
\label{ima}
\end{equation}
In this equation $x=m_{J/\psi}^2/W^2$ where $W$ denotes the total
CM energy of the $\gamma \gamma$ system, $m_{J/\psi}$ is the
mass of the $J/\psi$ meson, $\tdm Q_P / 2 \pm \tdm k$ denote the
transverse momenta of the exchanged gluons and $\tdm Q_P$ is the
transverse part of the momentum transfer. The role of the parameter
$s_0$ will be explained later.
The impact factor $\Phi_0(k^2, Q_P^2)$ describes the $\gamma J/\psi$
transition induced by two gluons and may be calculated
in QCD \cite{GINZBURG}.
The function $\Phi(x,\tdm k,\tdm Q_P)$ satisfies the non-forward BFKL equation
which may be represented symbolically as follows:
\begin{equation}
\Phi(\tdm Q_P) =\Phi_0(\tdm Q_P)+ {3\alpha_s(\mu^2)\over 2\pi^2}
{\cal K} (\tdm Q_P) \times \Phi(\tdm Q_P)
\label{bfkl}
\end{equation}
where the dependence of the impact factors $\Phi$'s and the BFKL kernel
${\cal K}$ on the transverse momenta and on the longitudinal momentum
fraction(s) is not indicated.
The scale of the QCD coupling $\alpha_s$ which appears in the impact
factors and in Eq.~(\ref{bfkl}) will be set to $\mu^2=k^2+Q_P^2/4 +m_c^2$
where $m_c$ denotes the mass of the charmed quark. The differential
cross-section is related in the following way to the amplitude~$A$:
\begin{equation}
{d \sigma \over dt} = {1\over 16 \pi} |A(W^2,t)|^2
\label{dsdt}
\end{equation}
It follows from the solution to the leading order BFKL equation
(\ref{bfkl}) (with the fixed coupling constant) that the differential
cross-section (\ref{dsdt}) rises for asymptotically large energies $W$
as $(W^2)^{2\lambda_P}$ with $\lambda_P$ given at the leading order by
the famous BFKL formula $\lambda_P = 12 \alpha_s\, \log(2) /\pi $.
However, this exponent is subject to very large NLO
corrections~\cite{BFKLNL}:
\begin{equation}
\lambda_P ^{(NLO)} \simeq \lambda_P (1 - 6.2 \alpha_s).
\label{lnlo}
\end{equation}
For all values of  $\alpha_s$, relevant for experiments, this
expansion is ill defined. It seems that the source of this problem can
be traced back to non-conservation of energy and momentum in the gluon
configurations resummed by the BFKL ladder. Although these effects
are formally subleading, they appear to be numerically large as it can be
seen e.g. in Eq.~(\ref{lnlo}). At present, it is not known how to
impose the exact phase space conditions on the multiple gluon
emissions which build up the BFKL ladder. One of the possible approximate
solutions to this problem was proposed in \cite{LDC} and developed in
\cite{KMS}. The proper kinematics is partially restored when one
requires that the virtualities of the gluons propagating along the ladder 
are dominated by the transverse momenta squared.
This modification, which we call kinematical or consistency constraint
does not introduce any effect at the leading order, but when taken
into account at the NLO accuracy, it exhausts about 70\% of
the exact NLO result. Moreover this constraint eliminates double
transverse logarithms in the DGLAP limits of the NLO BFKL kernel as
required by the renormalization group equations \cite{RESUM}. Thus,
besides equation (\ref{bfkl}) we shall consider a similar equation,
with the kinematical constraint imposed on the BFKL kernel. The
modified BFKL equation is more reliable and the cross-sections arising
from BFKL at LO are kept only as a reference.

Let as also mention how we treat the infra-red domain. We
introduce a parameter $s_0$ in the propagators of exchanged gluons.
This parameter can be viewed  as the effective representation of
the inverse of the colour confinement radius squared.
The parameter $s_0$ will be varied within the range
$(0.2 {\rm \; GeV})^2 < s_0<(0.4 {\rm \;GeV})^2$.
Sensitivity of the cross-section to its magnitude can serve as an
estimate of the sensitivity of the results to the contribution coming
from the infrared region. It should be noted that formula (\ref{ima})
gives finite result in the limit $s_0=0$. The results with finite
$s_0$ are however more realistic.

\noindent
\begin{figure}[hbpt]
\leavevmode
\begin{center}
\parbox{8cm}{
\epsfxsize = 7.5cm
\epsfysize = 7.5cm
\epsfbox[18 200 565 755]{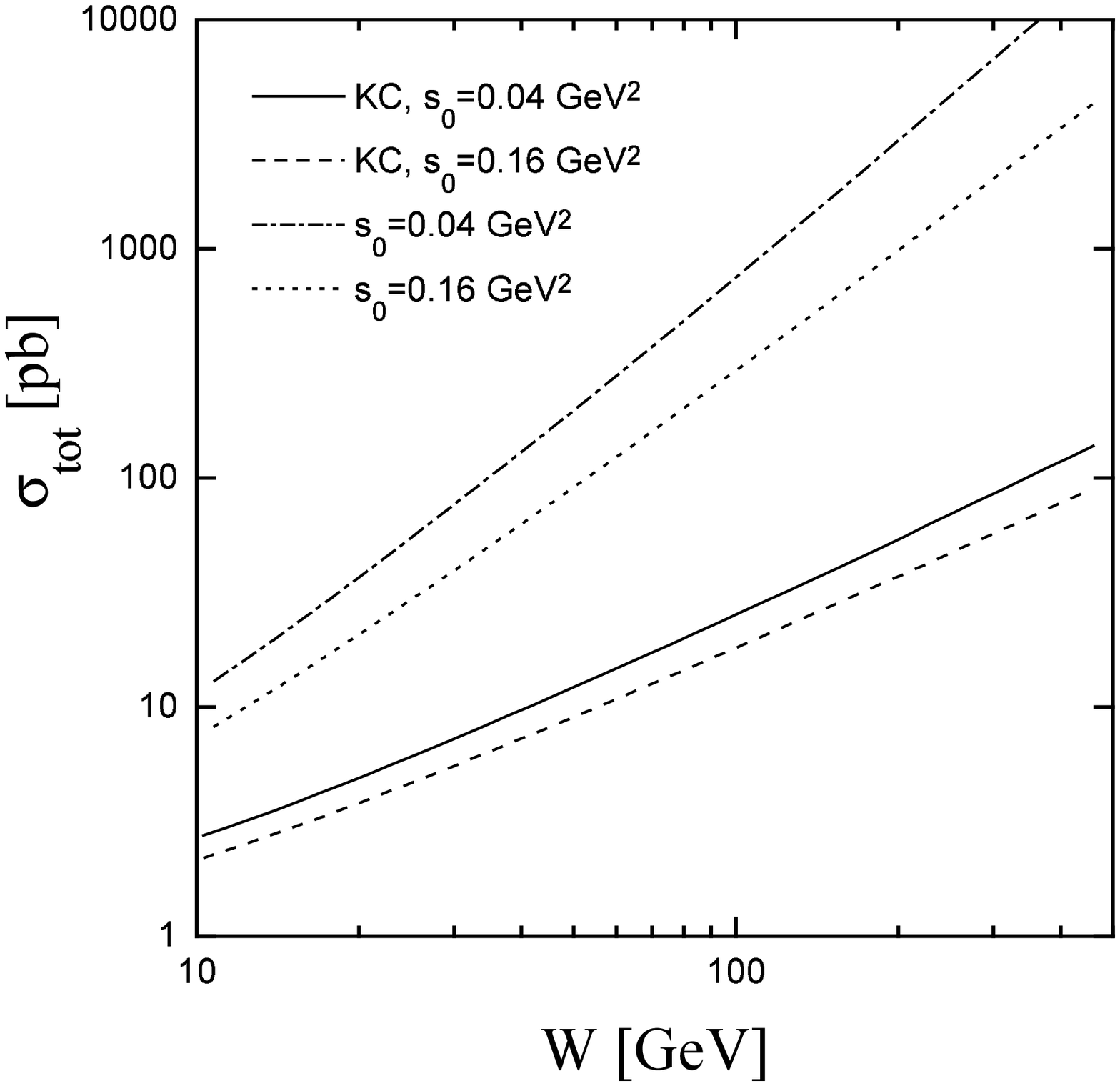}}
\end{center}
\caption{\small
Energy dependence of the cross-section for the process $\gamma\gamma
\rightarrow J/\psi J/\psi$. The two lower curves correspond to the
calculations based on the BFKL equation with kinematical constraint
and the values of $s_0$ equal to $0.04 {\rm \; GeV}^2$
(the continuous line)  and to $0.16 {\rm \; GeV}^2$ (dashed line).
The two upper curves correspond to the BFKL equation in the
leading logarithmic approximation with $s_0=0.04 {\rm \; GeV}^2$
(dash-dotted line) and $s_0=0.16{\rm \;GeV}^2$ (short-dashed line).}
\end{figure}

In Fig.~2 we show the obtained cross-section for the process
$\gamma\gamma \rightarrow J/\psi J/\psi$ plotted as a function of
the total CM energy $W$. We show results based on the BFKL equation in
the leading logarithmic approximation as well as those which include
the dominant non-leading effects. The calculations were performed for
 two values of the parameter $s_0$ i.e. $s_0=0.04 {\rm \; GeV}^2$
and $s_0=0.16 {\rm \; GeV}^2$.

Let us discuss the main features of the obtained results.
We see from Fig.~2 that the effect of the non-leading contributions is
very important and that they significantly reduce the magnitude of the
cross-section and slow down its increase with increasing CM energy~$W$.
The cross-section obtained from the BFKL equation with non-leading
corrections exhibits an approximate $(W^2)^{2\lambda_P}$
dependence. The parameter $\lambda_P$, which slowly varies with the
energy~$W$, takes the values $\lambda_P \sim 0.23 - 0.28$ within the
energy range $20{\rm \; GeV} < W < 500{\rm \; GeV}$ relevant for LEP2
and for possible TESLA measurements.
The cross-section calculated with the BFKL equation in the leading
logarithmic approximation gives a much stronger energy dependence of the
cross-section. The enhancement of the cross-section
is still appreciable after including the dominant non-leading
contribution which follows from the kinematical constraint. Thus while
in the Born approximation (i.e. for the elementary two gluon exchange
which gives energy independent cross-section) we get $\sigma_{tot}
\sim 1.9-2.6$~pb the cross-section calculated from the solution of the
BFKL equation with the non-leading effects taken into account can
reach the value 4~pb at $W=20 \gev$ and 26~pb for $W=100 \gev$ i.e.
for energies which should be easily accessible at TESLA.
The magnitude of the cross-section decreases with increasing magnitude
of the parameter $s_0$ which controls the contribution coming from the
infrared region. The energy dependence of the cross-section
is practically unaffected by the parameter $s_0$.

In our calculations we have assumed dominance of the imaginary part of
the production amplitude. The effect of the real part can be taken
into account by multiplying the cross-section by the correction factor
$1+tg^2(\pi\lambda_P/2)$ which for $\lambda_P \sim 0.25$ can introduce
an additional enhancement of about 20~\%.

In order to link our results to the $e^+e^-$ observables we consider
the most inclusive quantity which is the total cross-section
$\sigma_{tot} (e^+ e^- \to e^+ e^- J/\psi J/\psi)$. In fact, it is
convenient to impose additionally the anti-tagging condition. Taking
$\theta_{max}=30$~mrad and the minimal $\gamma\gamma$ energy to be
15~GeV we get for $\sigma_{tot} (e^+ e^- \to e^+ e^- J/\psi
J/\psi)$ the values of about 0.74~pb at $\sqrt{s}=500\gev$ (i.e. for
typical energies at TESLA). Thus, at TESLA for an $e^+e^-$ energy of
500~GeV and integrated luminosity of 100~fb$^{-1}$ we expect about
74~000 events, which is a large number, even if the acceptance for the
$J/\psi$ detection is low.

\begin{figure}[htb]
\leavevmode
\begin{center}
\parbox{12cm}{
\epsfxsize = 11cm
\epsfysize = 11cm
\epsfbox[0 130 600 700]{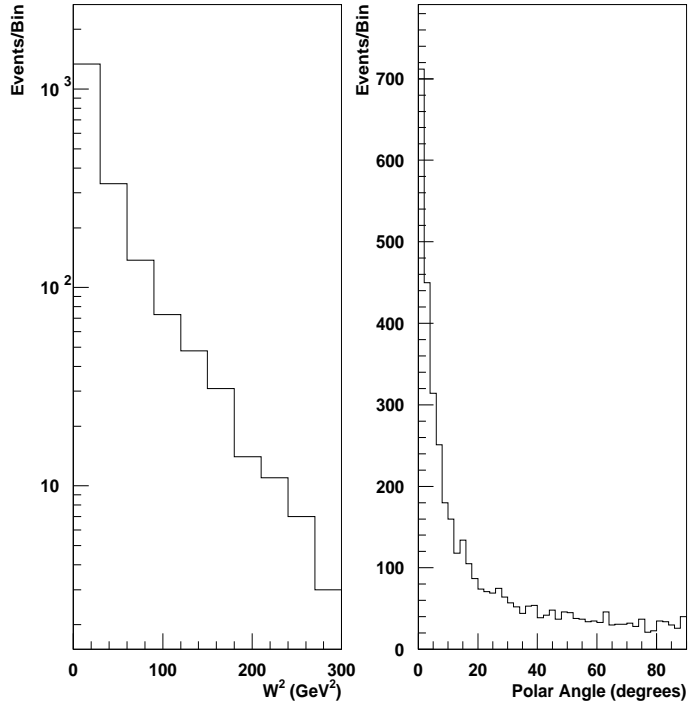}}
\caption{The $W$ distribution of the generated
events  and the distribution of the angle of the decay muons with 
respect to the beam direction.
}\label{fig4}
\end{center}
\end{figure}

The measurability of the $J/\psi$ mesons has been studied with the 
Monte Carlo generator PYTHIA, adapted for this purpose~\cite{SJOS}.
Here the reaction $\gamma\gamma \to J/\psi J/\psi$ is generated, with
zero transverse momentum transfer. The decays products of the 
$J/\psi$ will generally have small angles with respect to the 
beam particle direction, i.e. stay close to the beampipe, 
rendering experimental detection difficult. 
The most promising detectable decay channels are the leptonic decays
$J/\psi \to e^+e^-$ and $J/\psi \to \mu^+\mu^-$. 
Fig.~\ref{fig4} shows the $W$ distribution of the generated
events  and the distribution of the angle of the decay muons with 
respect to the beam direction.

A generic detector at the LC~\cite{CDR} has an acceptance for muons
and electrons starting from polar angles $\theta>$ 100 mrad.
If the background from the interaction region would be 
unexpectedly too large, the minimum 
reachable $\theta$ angle  will have a larger value,
e.g. 150 mrad.
Reaching  smaller angles is not impossible: the inner
mask, which occupies the detector region below 100 mrad, can be 
instrumented for detecting particles in the region  
$\theta = $ 20-25 to 100 mrad\cite{SITGES}. Since this region will suffer
from background electrons and photons, in case the $J/\psi$ decays into 
electrons, these electrons will have to have
at least 50 GeV to be detectable above this relatively uniform background.
In table~\ref{tab1} we show the acceptance (not taking into account
the magnetic field) for muons as function of the minimum angle 
$\theta_{min}$ and momentum $p_{tot}$ for detecting the particles,
for TESLA at $\sqrt{s}=500$ GeV. The detection efficiencies for the 
electrons are similar apart from the low angle region, where a 
high momentum of 50 GeV is always required. The corresponding values 
for the electrons with that additional 
requirement are given in brackets in the table.

\begin{table}[h]
\begin{center}
\begin{tabular}{|c|c|c|}
\hline
$p_{tot}$ & $\theta_{min}$ & Efficiency \\
\hline
1 GeV & 20 mrad & 73\% (48\%) \\
1 GeV & 100 mrad & 46\% \\
1 GeV & 150 mrad & 38\% \\
\hline
2.5 GeV & 20 mrad & 40\% (18\%) \\
2.5 GeV & 100 mrad & 17\% \\
2.5 GeV & 150 mrad & 10\% \\
\hline
5 GeV & 20 mrad & 25\% (8\%) \\
5 GeV & 100 mrad & 6\% \\
5 GeV & 150 mrad & 2\% \\
\hline
\end{tabular}
\end{center}
\caption{ Detection efficiency for $J/\psi$'s decaying into muons or electrons
(values between brackets take into account the 50 GeV 
momentum requirement for the 
electrons in the region of the mask). A $J/\psi$
is considered lost if one
of its decay particles is not within the acceptance region.}\label{tab1}
\end{table}

A realistic choice is 
$p_{tot}>2.5$ GeV, 
for which the efficiency is 40\% for the channel  $J/\psi \to \mu^+\mu^-$,
for the full angular region, and 17\% for the region $\theta_{min}>100$ 
mrad. Taking into account the branching ratio for 
$J/\psi \to \mu^+\mu^-$ ($\sim$ 6\%), the total number of detectable events 
in the 4-muon channel is 43 events
for 
$\theta_{min} > 20$ mrad and 100 fb$^{-1}$.
When also the electron channels are included, this increases to 
88 events.
In case    $\theta_{min} >100$ mrad, one expects 8 events (4-muons)
and 32 events (muon and electron channels).
These numbers
   show that this measurement will need high luminosity 
and an as large as possible  angular acceptance.

These efficiencies were estimated  for events with four-momentum
transfer $t = 0$.
The detectability increases with increasing $t$: at $|t|$
 = 4 GeV$^2$, the efficiency is a factor four better.
The efficiency decreases with increasing $W$ due to the 
kinematics: for $p_{tot}>2.5$ GeV, $\theta_{min} >100$ mrad, and
$W > 100$ GeV, the $J/\psi$ detection efficiency is reduced to 5\%.

In conclusion, the doubly exclusive $J/\psi$ production may be
successfully studied at TESLA providing important tests of
our understanding of the QCD pomeron dynamics.
A high integrated luminosity is however indispensable ($> 100$ fb$^{-1}$) 
and a good angular acceptance for the muons and/or electrons down
to small angles ($\theta_{min} \sim 20 $ mrad) mandatory.

\section*{Acknowledgments}
We are grateful to the Organizers for the interesting and stimulating
meeting. This research was partially supported
by the Polish State Committee for Scientific Research (KBN) grants
2~P03B~89~13, 2~P03B~084~14 and by the
EU Fourth Framework Programme 'Training and Mobility of Researchers', Network
'Quantum Chromodynamics and the Deep Structure of Elementary Particles',
contract FMRX--CT98--0194.

\end{document}